\documentstyle[prl,twocolumn,aps,psfig,epsfig,epsf,floats]{revtex}
\begin{document}
\draft
\title{Averaging rheological quantities in descriptions of soft glassy 
materials}
\author{Fran{\c c}ois Lequeux $^1$, Armand Ajdari $^2$}
\address{$^1$ 
Laboratoire de Physico-Chimie Macromol{\'e}culaire, UMR CNRS 7615\\
$^2$
 Laboratoire de Physico-Chimie Th{\'e}orique, Esa CNRS 7083\\
{\em both at :} ESPCI,
  10 rue Vauquelin, F-75231 Paris Cedex 05, France\\
submitted to Phys.Rev.E, Nov. 27, 2000}

\address{
\begin{minipage}{5.55in}
\begin{abstract}\hskip 0.15in
Many mean-field models have been introduced to
describe the mechanical behavior of glassy materials.
They often rely on averages performed over distributions
of elements or states. We here underline that averaging is a more intricate
procedure in mechanics than in more classical situations such
as phase transitions in magnetic systems.
This leads us to modify the predictions of the recently
proposed SGR model for soft glassy materials, for which we suggest that
the viscosity should diverge at the glass transition temperature $T_g$
with an exponential form $\eta \sim \exp(\frac{A}{T-T_g})$.
\end{abstract}
\end{minipage}
\vspace*{-0.3cm} 
}
\maketitle

\newcommand{\gd}{{\dot{\gamma}}}




\newpage
\section{Introduction}

Concentrated colloidal suspensions of soft particles
(emulsions, slurries)
tend to display glassy behavior when concentrated beyond
a threshold \cite{Mas1,Mas2,Ket1,Clo1,Der1}. 
Despite their great physical and chemical diversity,
these Soft Glassy Materials often share many mechanical features.
This has prompted theoreticians to construct generic models
\cite{Sol1,Heb1,Der2,Kur1,Ber1},
independent of the microscopic features of the systems, 
and often borrowed from 
statistical physics. The outcome of such a procedure
is often a description in terms of a large distribution
of relaxation processes acting simultaneously, a picture
supported by experimental facts for many glassy systems
\cite{Bou1,Got1}.

Any macroscopic mechanical response of the system 
then results from an averaging over these distributions.
In this paper we stress the importance of the averaging procedure
when one deals with mechanics. 
Averaging procedures have already been discussed
for non-mechanical situations in the non-ergodic
glassy phase
(e.g. statistical mechanics of spin glasses \cite{Bou1}).
However averaging of 
mechanical fields is a priori a different story \cite{Gol1}.
This also affects the description of the ergodic fluid phase
(e.g. the divergence of the viscosity when one approaches the
``glass transition'' from the fluid side).

To make our point we focus here
on the recently introduced  SGR model 
(Soft Glassy Rheology) \cite{Sol1} which incorporates
mechanics in a simple
picture of the glass transition introduced by Bouchaud.
The system is taken as a collection of blocks, that evolve
according to stochastic equations, driven by the imposed shear rate.
A control parameter of the model is the effective temperature that we
will denote here $T$. A transition occurs at $T_g$ below which the system
is no more ergodic (and the fluid displays a yield stress).
A remarkable feature of the model is that the viscosity
diverges at $2T_g$ (not $T_g$), so that in the interval
$T_g<T<2T_g$ the material is a power-law fluid.
This large domain of viscosity divergence, in the ergodic phase,
is a feature
absent in most other models for soft glassy materials \cite{Heb1,Der2,Ber1}.

After general comments about averaging procedures in mechanics (section II), 
we show below that this feature
is a consequence of the adopted averaging procedure (section III).
We propose an improvement of this procedure, that leads
to a Vogel-Fulcher divergence of the viscosity
at $T_g$: 
$\eta \sim \exp( \frac{A}{T-T_g})$.
We end by a short discussion and by drawing a
few research directions (section IV).

\section{Various averages for rheology/mechanics}

Let us start by clarifying the crucial role of the 
averaging procedure 
in rheology or mechanics, very different
from the one used in many classical areas,
such as magnetic systems, where the imposed field $H$
is well-defined , and the average magnetization is simply
the algebraic mean of the local magnetizations.
In rheological problems, the fields are
imposed through the boundaries of the system,
so that often
the shear rate and the shear stress play more or less
symmetric roles, which results in a wide variety of situations.

\begin{figure}
\centerline{\epsfig{figure=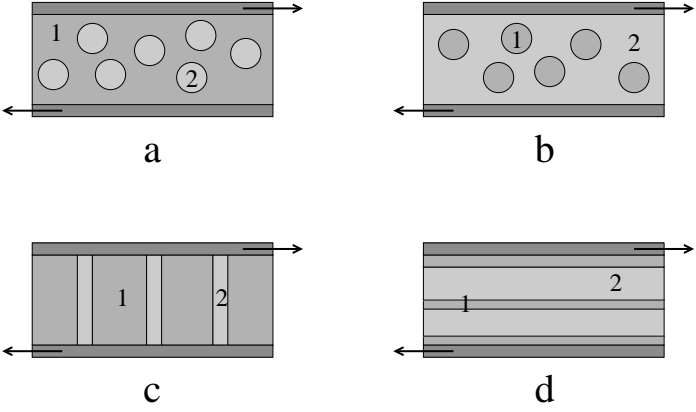}}
\vspace{.5cm}
\caption{Schematic pictures of : {\bf (a)} a dispersion
of fluid inclusions 2 in a highly viscous matrix 1,
{\bf (b)} a dispersion of highly viscous inclusions 1
in a fluid matrix 2, {\bf (c)} a geometry corresponding to averaging
at fixed strain rate ${\dot{\gamma}}$, which leads to a highly
viscous behavior as in (a), {\bf (d)} a geometry corresponding
to averaging at fixed stress with a qualitative behavior similar
to (b).
}\label{f_1}\end{figure}

To clarify the message, let us focus on the simple case
of a system built of elements
(blocks) of high viscosity $\eta_1$ and elements (blocks) of low viscosity
$\eta_2$ (Figs. 1a and 1b), and estimate the
average viscosity of the system through different averaging procedures.

A first average corresponds to the situation where the same shear rate 
$\dot{\gamma}$ is imposed to all elements, and one averages 
the resulting stresses, which leads to an average viscosity 

\begin{equation}
\left<\eta\right>_{\dot{\gamma}}= \phi_1 \eta_1 +(1-\phi_1)\eta_2
\end{equation}
where $\phi_1$
is the proportion of blocks of high viscosity. Strictly speaking
this procedure describes blocks arranged in layers disposed in parallel 
(Fig. 1c).
A second kind of average corresponds to submitting
all blocks to the same stress $\sigma$, and averaging the shear rates.
This results in a different
average viscosity 
\begin{equation}
\left<\eta\right>_{\sigma}= (\phi_1 \eta_1^{-1} +(1-\phi_1)
\eta_2^{-1})^{-1}
\end{equation}
which describes layers in series (Fig. 1d).

The behavior according to the first kind of average is dominated by the
highly viscous blocks: the average viscosity $\left<\eta\right>_{\dot{\gamma}}$
diverges if $\eta_1$ diverges, whatever the relative volume fractions. This 
is obviously a reasonable 
representation of a highly viscous medium with low-viscosity inclusions
($\phi_1$ large, Fig. 1a),
but a poor one for a system of highly viscous inclusions dispersed in a fluid
matrix of low viscosity $\eta_2$ (Fig. 1b).
On the contrary, the latter case ($\phi_1$ small, Fig. 1b)
is better represented by a fixed stress
average (Fig. 1d), 
as $\left<\eta\right>_{\sigma}$ is dominated by the low viscosity blocks.

Clearly none of these two simple averages properly describes all situations.
Consequently effective medium theories have tried to provide
better alternatives. Among those, let us quote the
Olroyd-Palierne formula \cite{Pal1}
for a matrix of viscosity $\eta_m$ with
various kind of dilute spherical inclusions {$i$} 
(volume fraction $\phi_i$, viscosity $\eta_i$).
This formula states that the effective viscosity of the resulting
composite fluid 
is (neglecting surface tension effects):
\begin{equation}
\left<\eta\right>_{OP}=\eta_m\frac{1+\frac{3}{2}\Sigma_i\phi_iH_i}
{1-\Sigma_i\phi_iH_i}
\end{equation}
with
\begin{equation}
H_i=\frac{2\eta_i-2\eta_m}{2\eta_i+3\eta_m}
\end{equation}
The above formula provides a convenient interpolation between the
two limiting physics mentioned above.
For a sparse
dispersion of highly viscous blocks in a less viscous matrix (Fig1.b),
it yields the Einstein formula 
$\left<\eta\right>_{OP}\simeq \eta_2 (1 +\frac{5}{2}\phi_1)$, and 
thus similar physics 
to the constant stress picture  
$\left<\eta\right>_{\sigma}\simeq \eta_2 (1+\phi_1)$ (Fig. 1d).
For a sparse dispersion of fluid inclusions in a highly viscous matrix 
(Fig. 1a),
$\left<\eta\right>_{OP} \simeq \eta_1 (1- \frac{5}{3}\phi_2)$, a result 
qualitatively
akin to that of the constant shear rate picture 
$\left<\eta\right>_{\dot{\gamma}}=\eta_1(1-\phi_2)$ (Fig. 1c).
Of course the Olroyd-Palierne formula is unable to describe accurately
the properties of intermediate mixtures (roughly $\phi$ in the range
$30-70 \%$) as it neglects spatial structures
that could result in percolation effects of great consequence
on the mechanical behavior.

\section{Consequences for the SGR model}

Implicitly the SGR model of Sollich et al. \cite{Sol1}
considers an ensemble of elements of viscosity distributed
along:
\begin{equation}
\eta(E)=\eta_0 \exp (E/T)
\end{equation}
where $E$ characterizes the activation barrier limiting the
yield of these blocks. The distribution of these
barriers is modeled by $n(E)\sim \exp (-E/T_g)$,
which naturally introduces $T_g$ as a (Glass) transition
temperature. Indeed, for $T>T_g$,
the states are populated according to Boltzmann statistics
at steady state $p(E)\sim n(E) \exp(E/T)$ which gives
$p(E)=\frac{T-T_g}{T_gT} \exp(- \frac{T-T_g}{T_gT}E)$,
which ceases to be normalizable as $T \rightarrow T_g^+$.

The average viscosity used in this model is the constant rate one,
which leads for $T>T_g$ to :
\begin{equation}
\left<\eta\right>_{\dot{\gamma}}=\int \!\!\!dE \, p(E)\, \eta(E)
= \int_0^{\infty}\! \!\!\!\! dE \frac{T-T_g}{TT_g} \, 
e^{- \frac{T-2T_g}{TT_g}E}
\end{equation}
which diverges for $T \rightarrow 2 T_g^+$ 
(i.e. not for $T\rightarrow T_g^+$). 
With this averaging procedure, the viscosity is thus infinite
in the range $T_g < T < 2T_g$, above the glass transition temperature.
The corresponding
mechanical behavior was found to 
be that of a power-law fluid with a temperature 
dependent exponent 
$\left<\sigma\right>\sim {\dot{\gamma}}^{\frac{T-T_g}{T_g}}$.

Let us now explain why, on physical grounds,
we expect this prediction to be incorrect,
and the viscosity of the system
to remain finite for $T_g<T<2 T_g$.
For example, focus on the $10 \%$ most viscous blocks
(i.e. those of viscosity larger than $\eta_c=\eta_0 \exp (E_c/T)$
with $E_c=-\frac{T_gT}{T-T_g}\log (0.1)$).
These few very viscous blocks are embedded 
in a matrix of viscosity
smaller than $\eta_c$, as 
all its blocks
have a viscosity weaker than $\eta_c$. 
The effective viscosity of the whole fluid,
composed of this matrix 
with sparse quasi-solid inclusions, is thus smaller
than $\eta_c$ times a geometric factor related to the $10 \%$ volume fraction
(e.g. $\eta < \frac{1.15}{0.9}\eta_c$ in the Olroyd-Palierne model), 
and the actual position of the inclusions. The viscosity is thus
finite for $T_g<T<2T_g$ and can not diverge faster 
than $\eta_c= \eta_0 \exp (\frac{T_g}{T-T_g}\log (10))$
as $T \rightarrow T_g^+$.

Switching to the other limit and using a fixed stress average,
leads to averaging the inverse of the viscosities (or the inverse
of the relaxation times). The continuous version
of Equation (2) yields an average viscosity:
\begin{equation}
\left< \eta \right>_{\sigma} = \eta_0 \frac{T}{T-T_g}
\end{equation}
which displays a weak (algebraic) divergence at $T_g$

We now propose a physically sounder 
recipe based on the Olroyd-Palierne formula
and a self-consistent approximation. We take a representative 
but small proportion $\phi$ of the blocks as ``inclusions'', while
the rest is the ``matrix''. As the matrix and the whole system are
similar, formula (3) gives an average viscosity 
$\left<\eta\right>_{OPSS}=\eta_m$:
\begin{equation}
\left<\eta\right>_{OPSS}=\left<\eta\right>_{OPSS}\frac{1+\frac{3}{2}
\Sigma_i\phi_i H_i}
{1- \Sigma_i\phi_i H_i}
\end{equation}
with
\begin{equation}
\Sigma_i\phi_i H_i= \phi \int_0^{\infty}\!\! dE \,\,p(E) 
\frac{2\eta(E)-2\left<\eta\right>_{OPSS}}{2\eta(E)
+3\left<\eta\right>_{OPSS}}
\end{equation}
The average viscosity $\left<\eta\right>_{OPSS}$ is thus implicitly given
by $\Sigma_i\phi_i H_i=0$ (a similar approximation
was used in another context \cite{Hem1}):
\begin{equation}
\int_0^{\infty}\!\! dE \,\, p(E) \frac{5\left<\eta\right>_{OPSS}}{2\eta(E)
+3\left<\eta\right>_{OPSS}}= 1
\end{equation}
Although the actual numerical factors (2 and 3) may be discussed,
we believe that this formula gives a physically correct average,
intermediate between the two limiting 
cases mentioned above: 
$\left<\eta\right>_{\dot{\gamma}} \ge \left<\eta\right>_{OPSS} \ge 
\left< \eta\right>_{\sigma}$, and 
with the appropriate behavior in the two limits.

Applying this to the SGR model of Sollich et al.
leads to a divergence of the viscosity at $T_g$.
Indeed if we write the average viscosity at temperature
$T$ in the form $\left<\eta\right>_{OPSS}=\frac{2}{3}\eta_0\exp(E^*/T)$,
then the implicit equation for $E^*$ is:
\begin{equation}
\int_0^{\infty}\!\! dE \,\,p(E) \frac{1}{\exp{\frac{E-E^*}{T}}+1}= 3/5
\end{equation}
Given that $\int_0^{\infty} dE\, p(E)=1$, this suggests 
$\int_{E^*}^{\infty}dE\, p(E) \sim 2/5$
and thus $\exp(-E^*\frac{T-T_g}{T_gT}) \sim 2/5$.
This corresponds to a viscosity diverging
for $T \rightarrow T_g^+$ as
$\sim \frac{2}{3}\eta_0 \exp( \frac{T_g}{T-T_g} \log(5/2) )$.
This divergence, obtained by the above rough analysis,
is confirmed by numerical inspection of (10) which gives
\begin{equation} 
\left<\eta\right>_{OPSS} 
\sim \eta_0 \exp( \frac{T_g}{T-T_g} \log(5/2) )
\end{equation}
with a prefactor $\simeq 0.72$ instead of $2/3$.
This exponential divergence is
consistent with our qualitative analysis below
equation (6).

Of course the $5/2$ is not to be taken too seriously,
but we think that a Vogel-Fulcher like divergence of the viscosity
at $T_g$
is a sound result, within the hypothesis of the SGR model.

\section{Discussion}

In conclusion, great care should be taken when choosing
an averaging procedure for the mechanical properties 
of a system close to a ``glass transition''. Fixed stress
and fixed shear rate averages are only extremes that limit
a large set of possibilities. A physically appealing
intermediate procedure for elements of distributed properties,
adapted from effective medium theory,
has been proposed in equation (10).

Focusing on the SGR model, we have shown that using different
averages seriously modifies the law characterizing the divergence of
the viscosity. Using the recipe of equation (10)
we have proposed that a Vogel-Fulcher divergence could quite naturally
arise. Revisiting the model, to incorporate the effects
of the inhomogeneity of the driving strain rate field $\dot{\gamma}$
 on the dynamics,
is a direction that we wish to explore.

It would also be interesting to analyze thoroughly
what kind of averages are implicitly made within the dynamical
rules used in other models for such systems \cite{Heb1,Der2,Ber1}. 
For example the simple
models of \cite{Heb1,Der2} predict an algebraic divergence
of the viscosity. A natural question is whether this is the result
of an implicit averaging of the inverse of the relaxation times
as in equation (7).

We have considered here the simplest (in principle)
situation of an ergodic system (i.e. above the effective 
glass transition temperature), where neglecting spatial structures
and adopting mean-field approaches is most likely
to apply. Analysis of non-ergodic situations is left for 
further work.
Eventually let us clarify that we have focused here on the 
sole viscosity to make our point clear, but that a more complete
description of rheological properties would have to deal with effective
theories for assemblies of visco-elastic (and possibly plastic)
elements \cite{Pal1}. Such effective theories most likely will have to deal
with the importance of spatial correlations
and with the role of percolating structures.\\

{\em Acknowledgments: we thank S. Obukhov for fruitful discussions
on related issues.}



\end{document}